\documentclass[twocolumn,secnumarabic,amssymb, nobibnotes, aps, pra, superscriptaddress]{revtex4-2}

\usepackage{CJKutf8}
\usepackage{amsmath}
\usepackage{bm}
\usepackage{graphicx}
\usepackage{float} 
\usepackage{subfigure} 
\usepackage{color}
\usepackage[mathlines]{lineno}
\usepackage[braket, qm]{qcircuit}
\usepackage{ulem}
\usepackage{svg}
\usepackage[ruled,noline,nofillcomment]{algorithm2e}
\usepackage{hyperref}

\begin{document}
\begin{CJK}{UTF8}{gbsn}
\title{Quantum coupon collector with mixed-state encoding}

\author{Jing-Peng Zhang}
\affiliation {Key Laboratory of Atomic and Subatomic Structure and Quantum Control (Ministry of Education), Guangdong Basic Research Center of Excellence for Structure and Fundamental Interactions of Matter, School of Physics, South China Normal University, Guangzhou 510006, China}

\author{Min-Quan He}
\email{hemq@hku.hk}
\affiliation{Guangdong-Hong Kong Joint Laboratory of Quantum Matter, Department of Physics, and HK Institute of Quantum Science \&  Technology, The University of Hong Kong, Pokfulam Road, Hong Kong, China}

\author{Dan-Bo Zhang}
\email{dbzhang@m.scnu.edu.cn}
\affiliation {Key Laboratory of Atomic and Subatomic Structure and Quantum Control (Ministry of Education), Guangdong Basic Research Center of Excellence for Structure and Fundamental Interactions of Matter, School of Physics, South China Normal University, Guangzhou 510006, China}

\affiliation{Guangdong Provincial Key Laboratory of Quantum Engineering and Quantum Materials,  Guangdong-Hong Kong Joint Laboratory of Quantum Matter, and Frontier Research Institute for Physics,\\  South China Normal University, Guangzhou 510006, China}

\begin{abstract}

 {The coupon collector is a prototypical model for evaluating the number of samples for identifying a set. By superposing all elements in the set as a pure quantum state, a quantum version of the coupon collector aims to learn the state, which is shown to reduce the sample complexity. Here we propose a quantum coupon collector by encoding the set into a mixed state, where the information of missing elements are labelled with Pauli strings. Remarkably, the encoded mixed state has no quantum entangled state and is easy to prepare.  With such mixed-state encoding, it can be efficient to learn the set by performing Bell measurements on two copies and then extracting the missing element by solving a series of equations obtained from the measurements. Our protocol further reduces the sample complexity from $O(n)$ in the case of pure-state encoding to $O(\log n)$ when the missing element is one, where $n$ is the number of elements in the set. The mixed-state encoding scheme provides a new avenue for quantum learning and enlarges the realm for exploring quantum advantages. }

\end{abstract}

\maketitle

\section{introduction}
\label{sec:introduction}
Computational Learning Theory was introduced in the 1980s by Leslie Valiant in his seminal paper “A Theory of the Learnable,”~\cite{valiant1984theory} which for the first time mathematically discussed and analyzed the concept of machine learning theory. In 1995, Bshouty and Jackson introduced quantum examples into the Dynamic Neural Field~(DNF) model~\cite{bshouty1995learning} and successfully proved that this model could learn quantum examples efficiently,  {which extends the learning theory in a quantum realm.}   {Remarkably,} within the framework of the Probably Approximately Correct (PAC) learning model \cite{haussler2018probably, valiant1984theory},  a quantum learner can demonstrate significant   {quantum} advantages over a classic learner~\cite{arunachalam2017survey}. Quantum-enhanced strategies for learning have shown advantages over traditional methods and have been extensively studied \cite{park2020geometry,gao2018experimental,servedio2001quantum,aaronson2018online,arunachalam2021two,grilo2019learning,huang2021information,bravyi2018quantum,zhong2020quantum,bremner2017achieving,gao2017quantum,bermejo2018architectures, zhu2022flexible}.

 {The coupon collector aims to collect a complete set of coupons.} {It can be formulated as to learn} the exact makeup of a $k$-sized unknown subset $S$ from a larger $n$-sized set $[n]$~\cite{atallah2009randomized,boneh1997coupon}. To acquire information about $S$, multiple copies are provided, each allowing the extraction of only one element. The  {goal} is to ascertain the minimal number of samples necessary to fully determine $S$,  {which is one of the major subjects investigated in }  learning theory~\cite{carbonell1983overview,mitchell1999machine,michie1995machine}, particularly in the context of  PAC learning \cite{haussler2018probably,valiant1984theory}. The classical approach  {for} solving this problem involves selecting an element randomly from each sample and iteratively learning an unseen element\cite{motwani1995randomized}.

A quantum version of the coupon collector problem was introduced by Arunachalam et al~\cite{arunachalam2020quantum}.  {By equal-weighting superposing all elements in a set into a pure quantum, the coupon collector can be turned into a quantum state learning problem.} For a small number of missing elements $m = n-k$ ,  {the} quantum strategy  {can} efficiently identify  {the} missing elements and significantly reduces the sample complexity from $O(n\log n)$ to $O(n)$. Subsequently, Zhou et al.~\cite{zhou2022experimental}  {experimentally} demonstrated a new quantum coupon collector protocol with a coherent-state encoding on a photonic quantum platform.  The existence of novel encoding methods for investigating the quantum collector problem is  {inspiring}.  {Particularly, the sample complexity depends heavily on the encoding scheme. The pure-state encoding has significant quantum advantages over the classical one in the sample complexity, at the price of comparatively complicated subroutines for state-preparation and quantum circuits for learning the state. While the coherent-state encoding simplifies the protocol, it has no quantum advantage when scaling up the size of the coupon collector}.  {It is wondered whether some alternative encoding schemes can give rise to advantages even over the pure-state encoding scheme and are also easy to implement.} 

 {In this paper, we propose a new encoding scheme for quantum coupon collector, where the missing elements are encoded as Pauli strings, mixed into a maximally mixed state. 
Such an encoding allows a $n$-sized set represented by $\log n$ qubits.  To learn the encoded state, we utilize Bell measurements on two copies of the encoded states and further locate the missing elements by solving equations obtained from the measurements. For one missing element, We show that the equations can be exactly solved with Gaussian elimination. The sample complexity turns to be $O(\log n)$, a requirement for obtaining a set of $\log n$ equations for decoding the Pauli string corresponding to the missing element. We point out that the mixed-state encoding scheme is not limited to coupon collector, but can provide an alternative avenue for studying the quantum version of learning, with the potential for exploring quantum advantages.}

 {The organization of the paper is as follows. In  Sec.~\ref{sec:mixed_state_encoding}, we review 
both classical and quantum coupon collector and propose the mixed-state encoding scheme. 
In Sec.~\ref{sec:measurement_and_post_processing}, we present the algorithm for learning the encoded mixed state with the Bell measurements and post-processing, with an illustrative example. Finally, discussions and conclusions are given in Sec.~\ref{sec:Conclusion and discussion}. }

\section{ {Encoding schemes in quantum coupon collector}}
\label{sec:mixed_state_encoding}
In this section, we introduce the encoding strategy of our protocol,  {which} encodes the coupon set state using a mixed quantum state. Before detailing our encoding strategy, we briefly review the classical encoding strategy~\cite{motwani1995randomized} and a representative quantum version~\cite{arunachalam2020quantum}. We then discuss our unique encoding strategy and its implementation.

In the classical coupon collection {problem}, the coupon state is represented as a set of coupons, where each coupon state is  {labeled as} a bit-string. Formally, the set is denoted as $S = \{ s_{i} \}_{i=1}^{n}$, where $n$ is the number of coupons to be collected, and each coupon state $s_{i}$ is a $b$-bit string with $b = \lceil \log_{2} n \rceil$. During the learning phase, each query randomly extracts one coupon state $s_{i}$ from the set $S$ with equal probability. The process aims to iteratively learn an unseen state $s_{k}$, with a probability of $(n-k)/n$ when $k$ coupons have been learned and $n-k$ coupons remain unseen. Therefore, the total number of samples required to learn the coupon set is $\sum_{k=0}^{n-1} n/(n-k)  {\approx} O(n \log n)$.

Arunachalam et al. proposed a quantum approach to the coupon collector problem that utilizes the power of quantum computation to reduce the sample complexity for a specific case~\cite{arunachalam2020quantum}.  {The coupon set $S$ is a subset of the whole set $[n]=\{1,2,3,...,n\}$. } In this quantum approach, both $S$ and  $[n]$ are encoded into equal-weighted superposition states  {$\ket{S}$} and $\ket{[n]}$,  {respectively}. The number of qubits required to encode the state $\ket{S}$ or $\ket{[n]}$ is $b =  \lceil \log_{2} n \rceil$. Measuring each copy of the coupon set state directly yields results similar to the classical case. However, unlike classical methods that measure the state $\ket{[n]}$ to identify elements, the quantum strategy employs amplitude amplification~\cite{brassard2002quantum} to extract information of the missing elements $\ket{\bar{S}}$ from the coupon state $\ket{[n]}$. With amplitude amplification, they demonstrate a reduced sample complexity, achieving $O(n)$ when the number of missing elements $m = n - k$ is small, compared to the classical $O(n \log n)$. This reduction is significant for this special case, though the implementation remains challenging due to the complexity of state preparation and {learning with amplitude amplification.}

{It is interesting to note that the state $S$, as an equal-weighting superposition of a subset of computations basis $\ket{i}$($i=1,2,...,n$), is typically highly entangled and thus quantum resource demanding. It is wondered whether a quantum strategy for coupon collector or other classical problems should demand less quantum resources. At the first stage, one can consider different encoding schemes when turning a classical problem into one that is suitable for quantum computing. }

{Here we consider an encoding scheme that the coupon set state can have no entanglement. Let us denote the $i$-th element in the coupon set $S$ as $s_i$. The missing elements are encoded as Pauli strings, }
\begin{equation}
s_i \rightarrow P_{s_i}=Z_1^{v_{i1}}Z_2^{v_{i2}}...Z_{b}^{v_{ib}},
\end{equation}
where $v_{i1}v_{i2}...v_{ib}$ is a binary representation of $s_i$, $I$ is the two-dimensional identity matrix  and $Z$ is the Pauli-Z matrix $\hat{\sigma}^{z}$. Then, those Pauli strings are mixed with a maximally mixed state. The general form of the coupon set state with $m$ coupons (or $k=n-m$ missing elements) can be written as,
\begin{equation}
\rho = 2^{-b} (I^{\otimes b}+\sum_{i=1}^{m}\frac{1}{m}\cdot P_{s_i}).
\end{equation}
Here $b = \lceil \log_{2} n \rceil$ is the number of qubits required to encode the state.  {The coefficient $\frac{1}{m}$ is introduced that the density matrix is positive semi-definite while some diagonal elements are zero.}

The mixed state is significantly easier to prepare than the equal-weighted quantum state, as it requires a shallower circuit depth as detailed later.  {Remarkably, the mixed state can be written as a mixing of computational basis and thus has no entanglement. Nevertheless, it still has some quantum correlations, e.g., quantum discord\cite{ollivier2001quantum,merali2011power}. It is expected that such quantum correlations, especially quantum discord in our case,  are necessary for realizing quantum advantage for the coupon collector problem, which leaves for further investigation. For our purpose, we consider a specific scenario with one missing element, which can be expressed as,}
\[\rho_1 = 2^{-b} (I^{\otimes b} + P).\] 
 {Here the subscript $1$ in $\rho_1$ means there is one missing element.}
Although our protocol focuses on the case where only one element is encoded into the coupon set state, this represents a typical case and provides insights for more general scenarios.

\begin{figure}[H]
    \centering
    \includegraphics[width=0.4\textwidth]{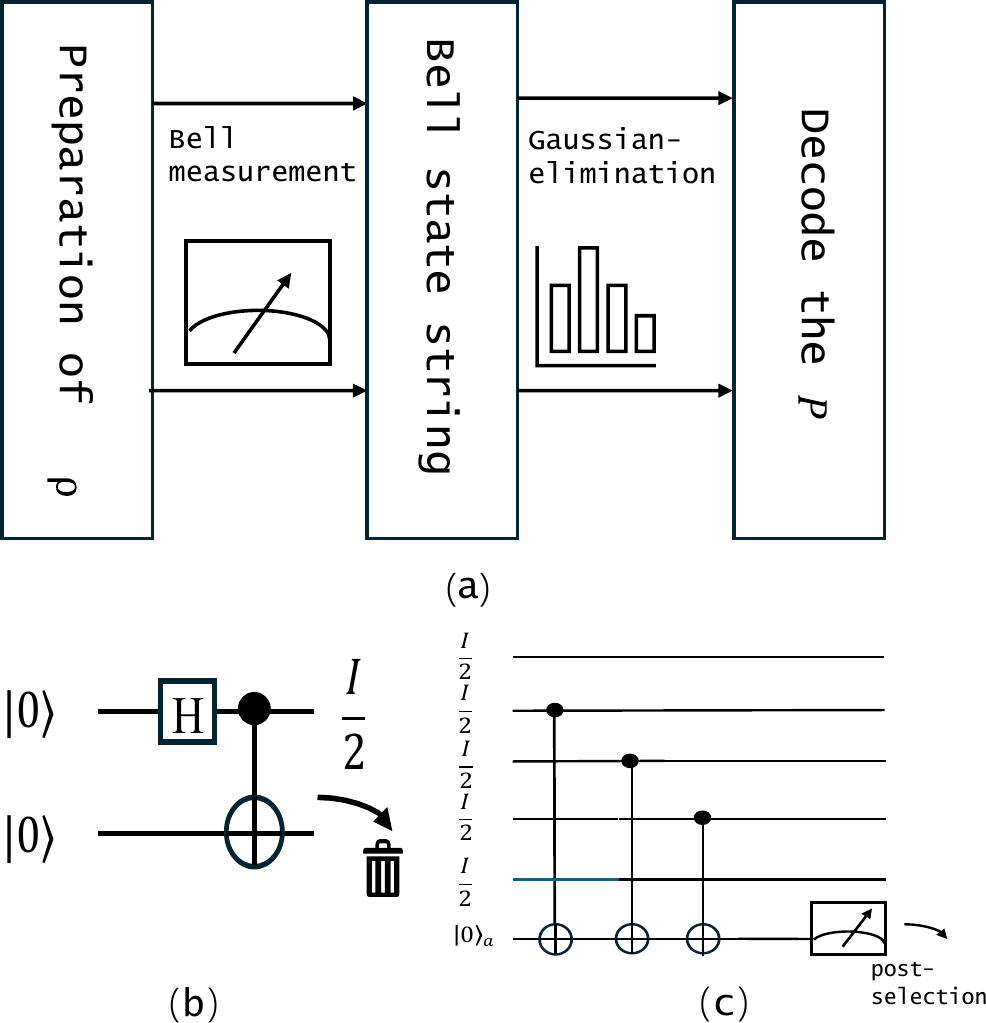}
    \caption{The framework of our protocol. (a): The state $\rho$ is prepared firstly, then with the BM, it is measured to Bell state string, in the end, the $P$ is decoded from the string by the  Gaussian elimination. (b): The circuit to prepare $\frac{I}{2}$. (c): The circuit to prepare \(\rho = 2^{-5} (I^{\otimes 5} + I \otimes Z \otimes Z\otimes Z\otimes I)\).  }
    \label{fig:framework}
\end{figure}

At the  {encoding} stage  {for} preparing the quantum coupon set state, we initially prepare a maximally mixed state \(I^{\otimes b}\). We then transform the maximum mixed state to the coupon set state \(\rho\). The  {procedure of state preparation} is outlined as follows:

\begin{itemize}
    \item[1.] \textbf{Preparation of the maximally mixed state.} For each qubit in the \(b\)-qubit register, perform a Hadamard gate followed by a CNOT gate with an auxiliary qubit. Trace out ( {discarding}) the auxiliary qubit to obtain the \(b\)-qubit maximally mixed state \(I^{\otimes b}\). The quantum circuit for these two steps for a single qubit is depicted in Fig.~\ref{fig:framework}b.
    \item[2.] \textbf{Transformation to the coupon set state}: Apply a CNOT gate for the qubit corresponding to the Pauli Z term in the Pauli string \(P\), where the control qubit is the register qubit and the target is the auxiliary qubit. Then measure the auxiliary qubit. If the result is \(\ket{0}\), the register qubits are in the desired state. A 5-qubit example with \(\rho = 2^{-5} (I^{\otimes 5} + I \otimes Z \otimes Z \otimes Z \otimes I)\) is shown in Fig.~\ref{fig:framework}.a.
\end{itemize}

In practical scenarios, the computational cost of decoding \(P\) can be substantial. To address this, we introduce the  Bell measurement~\cite{huang2022quantum} and the Gaussian elimination~\cite{lazard1983grobner, higham2011gaussian, sasaki1982efficient}, which help reduce computational costs. The outcomes of the  Bell measurement reveal the inherent properties of \(\rho\), which can be mapped to a coefficient matrix. The Gaussian elimination then   {exploits} these properties to simplify the coefficient matrix, aiding in the decoding of the correct \(P\). These methods reduce the sample complexity and post-processing time complexity to \(O(\log n)\) and \(O(\log^{3} n)\), respectively. The entire process is illustrated in Fig.~\ref{fig:framework}c, with detailed discussions of the decoding mechanisms provided in the subsequent section.

\section{ {Quantum state learning}}
\label{sec:measurement_and_post_processing}
In this section, we discuss the decoding strategy. Given a coupon set encoded into our mixed quantum state, the task of this stage is to decode the coupon information from the quantum state. Our decoding strategy adopts the  Bell measurement to sample informative bit-strings from the coupon quantum state, followed by classical post-processing to extract the Pauli string from these samples.

\subsection{Bell Measurement}
\label{subsec:entangling_bell_measurement}
Performing joint measurements on multiple coupons can be more efficient for learning a quantum state than measuring a single one independently\cite{carmeli2012informationally,uola2014joint,mcnulty2023estimating}. Here we adopt Bell measurements on two copies.
The resulting bit-string contains information about the measured quantum state and can be used to predict its properties, specifically  {$|\text{tr}(\hat{\sigma} \rho)|^{2}$}, where the observable operator $\hat{\sigma}$ is a Pauli string~\cite{huang2022quantum, cao2024accelerated}. Compared to direct measurement of quantum properties, this method reduces sample complexity. 

During the learning stage, every pair of qubits of two copies of states is projected under the Bell basis,
\begin{align}
      R_{i}^{(k)} \in \left\{ \ket{\Psi^+}\bra{\Psi^+}, \ket{\Psi^-}\bra{\Psi^-}, \ket{\Phi^+}\bra{\Phi^+}, \ket{\Phi^-}\bra{\Phi^-} \right\}
      \\
    \ket{\Psi^\pm}=\frac{1}{\sqrt{2}}(\ket{01}\pm\ket{10}), \ket{\Phi^\pm}=\frac{1}{\sqrt{2}}(\ket{00}\pm\ket{11}) \notag,
\end{align}
where $i$ is the $i$-th pair of qubits to be measured, $k$ denotes the $k$-th round of Bell measurements, and the results can be classically stored as a bit-string $[00, 01, 10, 11]$. For a single round measurement of a $b$-qubit quantum state, the  Bell measurement returns a $2b$ bit-string. With several rounds of measurements, the  {bit-strings} can be used to reconstruct the quantum state and classically predict the expectation value of the quantum state with an observable operator.
 
We first illustrate how the properties of a single qubit can be learned from the measured bit-strings.  {Note that the Bell states are eigenstates of $\hat{\sigma}\otimes \hat{\sigma}$ with eigenvalues $\pm 1$. The probability of getting a Bell state as an eigenstate with an eigenvalue $\pm1$ for $\hat{\sigma}\otimes \hat{\sigma}$ can be written as~\cite{huang2022quantum},}
\begin{equation}
    \begin{aligned}
        \text{Prob}(1) &=& \frac{1}{2} \text{tr}[(I \otimes I + \hat{\sigma} \otimes \hat{\sigma})(\rho \otimes \rho)],\\
        \text{Prob}(-1) &=& \frac{1}{2} \text{tr}[(I \otimes I - \hat{\sigma} \otimes \hat{\sigma})(\rho \otimes \rho)].
    \end{aligned}  
\end{equation}  
Then we have
\begin{equation}
    \begin{aligned}
        |\text{tr}(\hat{\sigma}\rho)|^2&= \text{tr}((\hat{\sigma} \otimes \hat{\sigma})(\rho \otimes \rho))\\ 
        &= \text{Prob}(1) - \text{Prob}(-1)\\
        &= \mathbb{E}[(\hat{\sigma} \otimes \hat{\sigma}) R],
    \end{aligned}
\end{equation}
where $\mathbb{E}$ indicates the expectation value and $R$ is the outcome Bell state after Bell measurement.  {For general multiple-qubit state with Pauli string $\hat{\sigma}$, estimation of      $|\text{tr}(\hat{\sigma}\rho )|^2$ turns to be~\cite{huang2022quantum},} 
\begin{equation}
    |\text{tr}(\hat{\sigma}\rho)|^2 = \mathbb{E}\left[\prod_{i=1}^b \text{tr}((\sigma_{i} \otimes \sigma_{i}) R_{i})\right],
    \label{eq:estimate_expectation}
\end{equation}
 {where $i$ is the qubit index.}

\subsection{Post-Processing} 
\label{subsec:post_processing}
 {The goal is to decode the Pauli string $P$  from results of Bell measurements.}
Recalling our encoded quantum state $\rho_1 = 2^{-b} (I^{\otimes b} + P)$, the estimated result $\text{tr}(\hat{\sigma}\rho_1) = 1$  {holds} only if the estimated Pauli string $\hat{\sigma}$ matches the encoded $P$ exactly; any other estimated Pauli string results in a zero expectation value.  {By natively trying all Pauli strings $\hat{\sigma}$, one can decode $P$. However, the time complexity grows exponentially with the number of qubits.}

To reduce the complexity of post-processing, it is necessary to exploit the condition $\text{tr}|(\hat{\sigma}\rho_1)|^2 = 1$ renders strong constraints on the relation between measurement results $\{ R_{i}^{(k)} \}$ and the candidate Pauli string $\hat{\sigma}$. Crucially, Eq.~\eqref{eq:estimate_expectation} can be simplified as,
\begin{equation}
|\text{tr}(\hat{\sigma}\rho)|^2 = \prod_{i=1}^b \text{tr}((\sigma_{i} \otimes \sigma_{i}) R_{i})=1, 
\label{eq:no_expectation}
\end{equation}
which can be considered as an equation by taking $\sigma_i=I,Z$ as variables. 
A series of equations can then be established from the results of Bell measurements and the candidate Pauli string. By solving the equations, the Pauli string $P$ can be decoded. As there are a number of $b=\log_2 n$ variables, the number of equations should be at least $\log_2 n$. Since  
each round of Bell measurements on two copies of $\rho_1$ will give an equation, the sample complexity turns to be $O(\log n)$. In the following, we solve the equations with Gaussian elimination, which takes a time complexity of $O(\log^3 n)$.

The original equations from Eq.~\eqref{eq:no_expectation} take symbols of $I$ and $Z$ as variables. 
Note that Bell states are eigenstates of $\sigma_{i} \otimes \sigma_{i}$ with 
eigenvalues $\pm 1$, namely $\text{tr}((\sigma_{i} \otimes \sigma_{i}) R_{i})=\pm1$. The choice of a positive and negative value is given in Table.~\eqref{table:correspondance}.
\begin{center}
    \begin{tabular}{|c|c|c|}
        \hline
        & \( I \otimes I \) & \( Z \otimes Z \) \\
        \hline
        \( \ket{\Phi^+}\bra{\Phi^+} \) & 1 & 1 \\
        \hline
        \( \ket{\Phi^-}\bra{\Phi^-} \) & 1 & 1 \\
        \hline
        \( \ket{\Psi^+}\bra{\Psi^+} \) & 1 & -1 \\
        \hline
        \( \ket{\Psi^-}\bra{\Psi^-} \) & 1 & -1 \\
        \hline
    \end{tabular} \label{table:correspondance}
\end{center}

 {The table shows that \( I \otimes I \) acts on any Bell state to yield a positive result; \( Z \otimes Z \) acts on the states \( \ket{\Phi^{\pm}}\bra{\Phi^{\pm}} \) to yield a positive one, and on the states \( \ket{\Psi^{\pm}}\bra{\Psi^{\pm}} \) to yield a minus one. From the table we can
find that $\text{tr}((\sigma_{i} \otimes \sigma_{i}) R_{i})=-1$ only when $\sigma_{i} = Z$ and $R_{i}^{(k)} = \ket{\Psi^{\pm}}\bra{\Psi^{\pm}}$ satisfy simultaneously. Then we can rewrite Eq.~\eqref{eq:no_expectation} as follows,} 
 {
\begin{equation}
    \prod_{i=1}^{b} (-1)^{(\sigma_{i} = Z \land R_{i}^{(k)} = \ket{\Psi^{\pm}}\bra{\Psi^{\pm}})}=1.\label{eq:single_component_transformation}
\end{equation}
}
Now the task is transformed to finding a Pauli string such that  {for each $k$-round of Bell measurements, the following equation should be satisfied,}
\begin{equation}
    \sum_{i=1}^{b} (\sigma_{i} = Z \land R_{i}^{(k)} = \ket{\Psi^{\pm}}\bra{\Psi^{\pm}}) \mod 2 = 0. 
    \label{eq:mod2_logical}
\end{equation}

 {Eq.~\eqref{eq:mod2_logical} can be further reformulated as a linear equation. The variables are defined as $x_i=0,1$ for $\sigma_{i} = I, Z$ respectively, while the coefficients are chosen as 
$g_{ki}=0$ for \( R_{i}^{(k)} = \ket{\Psi^{\pm}}\bra{\Psi^{\pm}} \) and $g_{ki}=1$ if \( R_{i}^{(k)} = \ket{\Phi^{\pm}}\bra{\Phi^{\pm}}\). Then Eq.~\eqref{eq:mod2_logical} takes the form of a linear equation, 
\begin{equation}
     \sum_{i=1}^{b} g_{ki} x_i  \mod 2 = 0. 
\end{equation}
}

 {For a total number of $t$ round Bell measurements, there are a number of $t$ equations. The system of linear equations can be neatly written in a matrix form,
\begin{equation}
    G\vec{x}\mod 2 = \vec{0},
    \label{eq:matrix_form}
\end{equation}
where $G$ is $k\times b$ matrix elements $G_{ki}=g_{ki}$, $\vec{x}=(x_1,x_1,...,x_{b})$ and $\vec{0}=(0,0,...,0)$ are $b$-dimensional vectors. } 

 {Eq.~\eqref{eq:matrix_form} is actually a system of linear equations over a finite field $Z_2$. To solve the equations, we refer to Gaussian elimination to further simplify the matrix $G$. The solution $\Vec{x}$ then can be obtained. The Pauli string then reads as, $P=Z_1^{x_1}Z_2^{x_2}...Z_{b}^{x_b}$. The whole procedure for decoding $P$ with Gaussian elimination as a subroutine is outlined in Algorithm.~\eqref{algorithm}.}

\SetCustomAlgoRuledWidth{8cm} 
\begin{algorithm}[H]
\KwData{coupon state $\rho$,  {results of Bell measurement} $R$, matrix $G$, simplified matrix $G'$}
\KwResult{Pauli string $P$}
  Prepare $\rho$ with the quantum circuit;\\
  Measure $\rho$ by BM, $\rho \rightarrow R$;\\
  Simplify $R$ in the degeneracy of Bell state;\\
  Construct coefficient matrix $G$ with $R$;\\
  $G'=\textbf{Gaussian elimination}(G)$;\\
  Decode $P$ from the simplified matrix $G'$;\\
 \caption{Decoding $P$ from $\rho$}
 \label{algorithm}
\end{algorithm}


 In our protocol, we need to construct an $\log n\times \log n$ size coefficient matrix to decode $P$.  {This requries $O(\log n)$ samples of coupon states in average.}
The time complexity of post-processing of Gaussian eliminations is $O(\log^3 n)$. 
\subsection{ {Illustrative example}} 
\label{subsec:Numerical demonstration}
For easier understanding, we provide a demonstration here. Initially, we select a Pauli string, $P=I\otimes Z\otimes Z\otimes Z\otimes I$. When $P$ is equal to $P=I\otimes Z\otimes Z\otimes Z\otimes I$,  The coupon state $\rho$ will be encoded as $\rho = 2^{-5} (I^{\otimes 5} + I \otimes Z \otimes Z \otimes Z \otimes I)$.

Following the preparation of the state $\rho$ (the quantum circuit is shown in Fig.~\ref{fig:framework}c), we create $6$ copies of it, and perform  Bell measurements on each set to acquire the measurement outcomes. 

From the measurement results, we arrive at the following coefficient matrix $G$. By Gaussian elimination, we can simplify $G$ to $G'$ as,
    \begin{equation}\notag
G
   =
    \begin{pmatrix}
0, &1, &1, &0, &1\\
 0, &1, &0, &1, &0\\
 1, &1, &0, &1, &0\\
 0, &0, &0, &0, &1\\
 0, &0, &0, &0, &0\\
 1, &1, &1, &0, &1\\
\end{pmatrix}
\rightarrow
G'=  \begin{pmatrix}
 1, &0, &0, &0, &0\\
 0, &1, &0, &1, &0\\
 0, &0, &1, &1, &0\\
 0, &0, &0, &0, &1\\
 0, &0, &0, &0, &0\\
 0, &0, &0, &0, &0\\
    \end{pmatrix}
\end{equation}

From the simplified coefficient matrix $G'$, we can get the equations and decode the value of $x_{b}$:
\begin{equation}\notag
    \begin{matrix}
         x_{1} \mod 2&=0\\
   x_{2} + x_{4} \mod 2&=0\\
   x_{3} + x_{4} \mod 2&=0\\
    x_{5} \mod 2 &=0\\
    \end{matrix}
      \longrightarrow
      \begin{matrix}
            x_{1}&=&0&\\
            x_{2}&=&1&\\ 
   x_{3}&=&1&\\
   
   x_{4}&=&1&\\
   x_{5}&=&0&\\ 
      \end{matrix}
\end{equation}

The value $x_{b}=0$ indicates that the corresponding Pauli operator is the Pauli operator $I$, while $x_{b}=1$ signifies that the Pauli operator is the matrix $Z$. Consequently, we can deduce the Pauli string that  $P=I\otimes Z\otimes Z\otimes Z\otimes I$.

\section{Conclusion and discussion}
\label{sec:Conclusion and discussion}
  {To summary,} we  {have introduced} a protocol for the quantum coupon collector problem by encoding the coupon state $\rho$ as a mixed state. Having Utilized  Bell measurements and Gaussian elimination, {for the case of one missing element,} the sample complexity and time complexity of post-processing have turned to be  $O(\log n)$ and $O(\log ^3 n)$, respectively,  {which has shown a significant reducing of sample complexity compared to the original pure-state encoding approach.}

The quantum coupon collector problem centers on determining the minimum number of samples required to learn an unknown set $S$. This problem is intrinsically connected to learning theory. As noted in reference \cite{arunachalam2020quantum}, quantum coupons show promise in proper PAC learning. We  {have pointed out} that our new protocol  {may} invigorate this field and anticipate that it may find additional applications in areas such as quantum communication and quantum cryptography. 

\begin{acknowledgments}
This work was supported by the National Natural Science Foundation of China (Grant No.12375013) and the Guangdong Basic and Applied Basic Research Fund (Grant No.2023A1515011460).
\end{acknowledgments}

\normalem

\end{CJK}

\begin{thebibliography}{100}

	\bibitem{valiant1984theory} Leslie~G Valiant.
\newblock A theory of the learnable.
\newblock {\em Communications of the ACM}, 27(11):1134--1142, 1984.

	\bibitem{bshouty1995learning} Nader~H Bshouty and Jeffrey~C Jackson.
\newblock Learning dnf over the uniform distribution using a quantum example oracle.
\newblock In {\em Proceedings of the eighth annual conference on Computational learning theory}, pages 118--127, 1995.

	\bibitem{haussler2018probably} David Haussler and Manfred Warmuth.
\newblock The probably approximately correct (pac) and other learning models.
\newblock {\em The Mathematics of Generalization}, pages 17--36, 2018.

	\bibitem{arunachalam2017survey} Srinivasan Arunachalam and Ronald de~Wolf.
\newblock A survey of quantum learning theory.
\newblock {\em arXiv preprint arXiv:1701.06806}, 2017.

	\bibitem{park2020geometry} Chae-Yeun Park and Michael~J Kastoryano.
\newblock Geometry of learning neural quantum states.
\newblock {\em Physical Review Research}, 2(2):023232, 2020.

	\bibitem{gao2018experimental} Jun Gao, Lu-Feng Qiao, Zhi-Qiang Jiao, Yue-Chi Ma, Cheng-Qiu Hu, Ruo-Jing Ren, Ai-Lin Yang, Hao Tang, Man-Hong Yung, and Xian-Min Jin.
\newblock Experimental machine learning of quantum states.
\newblock {\em Physical review letters}, 120(24):240501, 2018.

	\bibitem{servedio2001quantum} Rocco~A Servedio and Steven~J Gortler.
\newblock Quantum versus classical learnability.
\newblock In {\em Proceedings 16th Annual IEEE Conference on Computational Complexity}, pages 138--148. IEEE, 2001.

	\bibitem{aaronson2018online} Scott Aaronson, Xinyi Chen, Elad Hazan, Satyen Kale, and Ashwin Nayak.
\newblock Online learning of quantum states.
\newblock {\em Advances in neural information processing systems}, 31, 2018.

	\bibitem{arunachalam2021two} Srinivasan Arunachalam, Sourav Chakraborty, Troy Lee, Manaswi Paraashar, and Ronald De~Wolf.
\newblock Two new results about quantum exact learning.
\newblock {\em Quantum}, 5:587, 2021.

	\bibitem{grilo2019learning} Alex~B Grilo, Iordanis Kerenidis, and Timo Zijlstra.
\newblock Learning-with-errors problem is easy with quantum samples.
\newblock {\em Physical Review A}, 99(3):032314, 2019.

	\bibitem{huang2021information} Hsin-Yuan Huang, Richard Kueng, and John Preskill.
\newblock Information-theoretic bounds on quantum advantage in machine learning.
\newblock {\em Physical Review Letters}, 126(19):190505, 2021.

	\bibitem{bravyi2018quantum} Sergey Bravyi, David Gosset, and Robert K{\"o}nig.
\newblock Quantum advantage with shallow circuits.
\newblock {\em Science}, 362(6412):308--311, 2018.

	\bibitem{zhong2020quantum} Han-Sen Zhong, Hui Wang, Yu-Hao Deng, Ming-Cheng Chen, Li-Chao Peng, Yi-Han Luo, Jian Qin, Dian Wu, Xing Ding, Yi~Hu, et~al.
\newblock Quantum computational advantage using photons.
\newblock {\em Science}, 370(6523):1460--1463, 2020.

	\bibitem{bremner2017achieving} Michael~J Bremner, Ashley Montanaro, and Dan~J Shepherd.
\newblock Achieving quantum supremacy with sparse and noisy commuting quantum computations.
\newblock {\em Quantum}, 1:8, 2017.

	\bibitem{gao2017quantum} Xun Gao, Sheng-Tao Wang, and L-M Duan.
\newblock Quantum supremacy for simulating a translation-invariant ising spin model.
\newblock {\em Physical review letters}, 118(4):040502, 2017.

	\bibitem{bermejo2018architectures} Juan Bermejo-Vega, Dominik Hangleiter, Martin Schwarz, Robert Raussendorf, and Jens Eisert.
\newblock Architectures for quantum simulation showing a quantum speedup.
\newblock {\em Physical Review X}, 8(2):021010, 2018.

	\bibitem{zhu2022flexible} Yan Zhu, Ya-Dong Wu, Ge~Bai, Dong-Sheng Wang, Yuexuan Wang, and Giulio Chiribella.
\newblock Flexible learning of quantum states with generative query neural networks.
\newblock {\em Nature communications}, 13(1):6222, 2022.

	\bibitem{atallah2009randomized} Mikhail~J Atallah and Marina Blanton.
\newblock Randomized algorithms.
\newblock In {\em Algorithms and Theory of Computation Handbook, Volume 1}, pages 335--358. Chapman and Hall/CRC, 2009.

	\bibitem{boneh1997coupon} Arnon Boneh and Micha Hofri.
\newblock The coupon-collector problem revisited鈥攁 survey of engineering problems and computational methods.
\newblock {\em Stochastic Models}, 13(1):39--66, 1997.

	\bibitem{carbonell1983overview} Jaime~G Carbonell, Ryszard~S Michalski, and Tom~M Mitchell.
\newblock An overview of machine learning.
\newblock {\em Machine learning}, pages 3--23, 1983.

	\bibitem{mitchell1999machine} Tom~M Mitchell.
\newblock Machine learning and data mining.
\newblock {\em Communications of the ACM}, 42(11):30--36, 1999.

	\bibitem{michie1995machine} Donald Michie, David~J Spiegelhalter, Charles~C Taylor, and John Campbell.
\newblock {\em Machine learning, neural and statistical classification}.
\newblock Ellis Horwood, 1995.

	\bibitem{motwani1995randomized} Rajeev Motwani and Prabhakar Raghavan.
\newblock {\em Randomized algorithms}.
\newblock Cambridge university press, 1995.

	\bibitem{arunachalam2020quantum} Srinivasan Arunachalam, Aleksandrs Belovs, Andrew~M Childs, Robin Kothari, Ansis Rosmanis, and Ronald De~Wolf.
\newblock Quantum coupon collector.
\newblock {\em arXiv preprint arXiv:2002.07688}, 2020.

	\bibitem{zhou2022experimental} Min-Gang Zhou, Xiao-Yu Cao, Yu-Shuo Lu, Yang Wang, Yu~Bao, Zhao-Ying Jia, Yao Fu, Hua-Lei Yin, and Zeng-Bing Chen.
\newblock Experimental quantum advantage with quantum coupon collector.
\newblock {\em Research}, 2022, 2022.

	\bibitem{brassard2002quantum} Gilles Brassard, Peter Hoyer, Michele Mosca, and Alain Tapp.
\newblock Quantum amplitude amplification and estimation.
\newblock {\em Contemporary Mathematics}, 305:53--74, 2002.

	\bibitem{ollivier2001quantum} Harold Ollivier and Wojciech~H Zurek.
\newblock Quantum discord: a measure of the quantumness of correlations.
\newblock {\em Physical review letters}, 88(1):017901, 2001.

	\bibitem{merali2011power} Zeeya Merali.
\newblock The power of discord: physicists have always thought quantum computing is hard because quantum states are incredibly fragile. but could noise and messiness actually help things along?
\newblock {\em Nature}, 474(7349):24--27, 2011.

	\bibitem{huang2022quantum} Hsin-Yuan Huang, Michael Broughton, Jordan Cotler, Sitan Chen, Jerry Li, Masoud Mohseni, Hartmut Neven, Ryan Babbush, Richard Kueng, John Preskill, et~al.
\newblock Quantum advantage in learning from experiments.
\newblock {\em Science}, 376(6598):1182--1186, 2022.

	\bibitem{lazard1983grobner} Daniel Lazard.
\newblock Gr{\"o}bner bases, gaussian elimination and resolution of systems of algebraic equations.
\newblock In {\em European Conference on Computer Algebra}, pages 146--156. Springer, 1983.

	\bibitem{higham2011gaussian} Nicholas~J Higham.
\newblock Gaussian elimination.
\newblock {\em Wiley Interdisciplinary Reviews: Computational Statistics}, 3(3):230--238, 2011.

	\bibitem{sasaki1982efficient} Tateaki Sasaki and Hirokazu Murao.
\newblock Efficient gaussian elimination method for symbolic determinants and linear systems.
\newblock {\em ACM Transactions on Mathematical Software (TOMS)}, 8(3):277--289, 1982.

	\bibitem{carmeli2012informationally} Claudio Carmeli, Teiko Heinosaari, and Alessandro Toigo.
\newblock Informationally complete joint measurements on finite quantum systems.
\newblock {\em Physical Review A}, 85(1):012109, 2012.

	\bibitem{uola2014joint} Roope Uola, Tobias Moroder, and Otfried G{\"u}hne.
\newblock Joint measurability of generalized measurements implies classicality.
\newblock {\em Physical review letters}, 113(16):160403, 2014.

	\bibitem{mcnulty2023estimating} Daniel McNulty, Filip~B Maciejewski, and Micha{\l} Oszmaniec.
\newblock Estimating quantum hamiltonians via joint measurements of noisy noncommuting observables.
\newblock {\em Physical Review Letters}, 130(10):100801, 2023.

	\bibitem{cao2024accelerated} Chenfeng Cao, Hiroshi Yano, and Yuya~O. Nakagawa.
\newblock Accelerated variational quantum eigensolver with joint bell measurement.
\newblock {\em Phys. Rev. Res.}, 6:013205, Feb 2024.

	\bibitem{micciancio2009lattice} Daniele Micciancio and Oded Regev.
\newblock Lattice-based cryptography.
\newblock In {\em Post-quantum cryptography}, pages 147--191. Springer, 2009.

	\bibitem{nejatollahi2019post} Hamid Nejatollahi, Nikil Dutt, Sandip Ray, Francesco Regazzoni, Indranil Banerjee, and Rosario Cammarota.
\newblock Post-quantum lattice-based cryptography implementations: A survey.
\newblock {\em ACM Computing Surveys (CSUR)}, 51(6):1--41, 2019.

	\bibitem{aaronson2018shadow} Scott Aaronson.
\newblock Shadow tomography of quantum states.
\newblock In {\em Proceedings of the 50th annual ACM SIGACT symposium on theory of computing}, pages 325--338, 2018.

	\bibitem{aaronson2019gentle} Scott Aaronson and Guy~N Rothblum.
\newblock Gentle measurement of quantum states and differential privacy.
\newblock In {\em Proceedings of the 51st Annual ACM SIGACT Symposium on Theory of Computing}, pages 322--333, 2019.

	\bibitem{buadescu2021improved} Costin B{\u{a}}descu and Ryan O'Donnell.
\newblock Improved quantum data analysis.
\newblock In {\em Proceedings of the 53rd Annual ACM SIGACT Symposium on Theory of Computing}, pages 1398--1411, 2021.


\end{thebibliography}
\end{document}